\documentclass{gji}
\usepackage{timet,color}
\usepackage[urlcolor=blue,citecolor=black,linkcolor=black]{hyperref}
\usepackage{amsmath}
\title[PINNs for viscoacoustic wave propagation]
{Physics-informed neural networks for viscoacoustic wave propagation: forward modelling, inversion and discretization sensitivity}

\author[Liang et al.]
{
Chaohua Liang$^{1}$,
Xingliang Peng$^{2}$,
Jun Matsushima$^{1}$\thanks{Corresponding author: \texttt{jun-matsushima@edu.k.u-tokyo.ac.jp}}\\
$^{1}$ The University of Tokyo\\
Graduate School of Frontier Sciences\\
Kashiwa, Japan\\
$^{2}$ College of Water Conservancy, Jiangxi University of Water Resources and Electric Power\\
Nanchang 330099, China
}

\date{}
\volume{Preprint submitted to arXiv}
\pubyear{\the\year}
\pagerange{}

\usepackage{graphicx}

\begin{document}

\label{firstpage}

\maketitle

\begin{abstract}
Seismic wave forward and inverse modeling not only provide the theoretical foundation for understanding the Earth’s complex deep structures, but also constitute core techniques for subsurface resource exploration and geological hazard assessment. Traditional numerical approaches, such as the finite-difference method (FDM) and finite-element method (FEM), typically rely on grid-based discretization and iterative forward simulations when solving inverse problems. Although deep learning has shown considerable potential in seismic applications, its effectiveness is often limited by the requirement for large volumes of labelled data and the insufficient incorporation of physical constraints.
In this study, we propose a unified physics-informed neural network (PINN) framework for both forward modeling and parameter inversion of viscoacoustic wave propagation. The proposed framework is validated through comparative numerical experiments. For forward modeling, the network accurately reproduces wavefields, amplitude attenuation, and phase characteristics across different velocity models. For inversion, we evaluate the reconstruction of velocity and attenuation models from temporally sparse observations and random initial guesses. The results indicate that the proposed PINN approach achieves stable and acceptable accuracy when benchmarked against reference solutions, while exhibiting reduced sensitivity to spatial discretization compared with grid-based numerical methods.
More importantly, by embedding a viscoacoustic wave equation into the learning process, the framework provides a physically consistent representation of intrinsic attenuation effects beyond purely acoustic approximations. In addition, the capability of simultaneously recovering velocity and attenuation parameters under sparse sampling conditions highlights the robustness and flexibility of the proposed method. These results demonstrate the feasibility of the method and highlight its potential for high-resolution seismic exploration.

\end{abstract}

\begin{keywords}
Physics-Informed Neural Networks (PINNs), Viscoacoustic wave, Forward and inverse modeling, Attenuation and dispersion, Seismic exploration.
\end{keywords}

\section{Introduction}

Accurate seismic imaging is essential for resolving subsurface geological structures, particularly in the exploration and assessment of energy resources such as methane hydrates, commonly referred to as “combustible ice” owing to their high energy density and widespread occurrence. Over the past few decades, numerical methods such as the finite difference method (FDM) and finite element method (FEM) have been firmly established as the standard tools for seismic forward modeling and inverse analysis \cite{Liu2011ScalarWave,Nogueira2021RTM,Pled2022PMLReview}. These grid-based approaches typically discretize the domain into meshes to approximate the partial differential equations (PDEs). Although conventional approaches, such as FDM and full waveform inversion (FWI), can achieve high accuracy in solving partial differential equations, they rely on repeated high-fidelity forward simulations during inversion, which leads to substantial computational cost. What’s more, the misfit function of FWI is prone to converging toward local minima rather than the global optimum \cite{Virieux2009FWI}.

With the rapid evolution of deep learning (DL), together with the exponential increase in GPU-accelerated computing power, data-driven paradigms have witnessed explosive growth in the field of seismology. The enhanced hardware capabilities have allowed researchers to train sophisticated neural networks on massive seismic datasets, establishing complex non-linear mappings between measurements and subsurface properties. For example, Wu and Lin \cite{Wu2020InversionNet} proposed InversionNet, a framework that employs a deep convolutional encoder-decoder architecture to directly learn the non-linear mapping from seismic data to subsurface velocity structures. Similarly, Recurrent Neural Networks (RNNs) have been successfully applied to seismic impedance inversion \cite{Alfarraj2019Impedance}, leveraging the temporal correlations within seismic traces. Furthermore, Generative Adversarial Networks (GANs) have shown remarkable performance in seismic data reconstruction \cite{Li2025GAN}. However, these supervised methods require lots of labeled data, which is difficult to obtain in real-world seismic exploration. Besides, these models are essentially black-box mappers, therefore may generate physically unreasonable results.

Recently, Physics-informed neural networks (PINNs) by Raissi \cite{Raissi2019PINN} have received widespread attention as a new scientific computation algorithm. PINNs explicitly embed the governing Partial Differential Equations (PDEs) into the loss function as a regularization mechanism, which provides new ideas for addressing problems faced in conventional methods. Specifically, in inverse problems, PINNs integrate the forward and inverse processes into a unified framework. By treating the unknown parameters in PDEs as trainable variables, PINNs leverage automatic differentiation to update these parameters simultaneously with the outputs of the forward process. PINNs is now applied to miscellaneous scientific domains including fluid mechanics \cite{Raissi2019PINN,Raissi2020Hidden}, geophysics \cite{RashtBehesht2021PINNWave,Ren2022SeismicNet}, solid mechanics \cite{Haghighat2021Solid}, heat transfer \cite{Zhao2025Review} and biomedical imaging \cite{Wong2025Heart}.

However, vanilla PINNs face certain challenges in seismic exploration. Studies indicate that vanilla PINNs suffer from point-source singularity in solving wave equations. To address this problem, a hybrid initialization strategy that utilizes the solutions generated by the FDM method at early time steps as initial constraints is proposed \cite{Alkhalifah2021Helmholtz}. What’s more, while Physics-Informed Neural Networks (PINNs) have recently emerged as a powerful tool for solving seismic wave equations, the majority of existing studies restrict their scope to ideal acoustic or elastic media \cite{RashtBehesht2021PINNWave,Ren2022SeismicNet,Zou2024VelocityPINN}, explicitly neglecting the intrinsic attenuation properties of the subsurface. However, in realistic geological formations, the viscosity of rocks and the presence of fluids inevitably cause the conversion of mechanical energy into heat. This phenomenon, known as seismic absorption, induces velocity dispersion and amplitude dissipation, which significantly alter the waveform characteristics \cite{Malinowski2011Attenuation}. Furthermore, their application to inverse problems—particularly in complex viscoacoustic media—remains noticeably underexplored. To address these limitations, we establish a physics-informed architecture specifically designed to capture these complex viscoacoustic behaviors and is able to run forward and inverse process simultaneously.

In this paper, we propose a unified physics-informed neural network framework for viscoacoustic wave propagation, capable of both forward modeling and physical parameter inversion. To avoid point-source singularities, early-time wavefields are incorporated as initial constraints, and open boundary conditions are employed to suppress artificial boundary reflections.
The main contributions of this work are threefold. First, we demonstrate that the proposed framework accurately models viscoacoustic wavefields, capturing both attenuation and phase characteristics in homogeneous and layered velocity models. Second, we show that the PINN formulation remains stable under coarse discretization and sparse sampling conditions, exhibiting a fundamentally different degradation behavior compared to finite-difference solvers. Third, we introduce a simultaneous forward–inverse strategy that enables robust recovery of both velocity and absorption parameters from random initial models, highlighting the framework’s ability to solve the inverse problem by treating physical parameters as learnable variables.

\section{Problem Formulation}
In this paper, our goal is to investigate the ability of PINNs for viscoacoustic propagation in 2D homogeneous media. The governing equations based on the Maxwell rheological model for propagating pressure in the time-space domain are given by \cite{Deng2007Migration}:
\begin{equation}
\frac{1}{v^2} \frac{\partial^2 P}{\partial t^2}
- \nabla^2 P
+ \frac{g}{v} \frac{\partial P}{\partial t}
= S(x_s, t)
\label{eq:viscoacoustic_wave}
\end{equation}

Where \( P \), \( v \), and \( t \) denote the pressure, velocity, and time, respectively. \( S \) represents the source term. Consistent with standard seismic modeling practices, a Ricker wavelet is employed as the source function in this study. \( g \) is the absorption coefficient, given by:
\begin{equation}
g = \frac{2\pi f_0}{vQ}
\label{eq:absorption}
\end{equation}
Where \( f_0 \) and \( Q \) denote the reference frequency and the quality factor, respectively. The quality factor \( Q \) is analytically derived from Li's empirical formula:
\begin{equation}
Q = 3.516 \times 10^{-6} \times v^{2.2}
\label{eq:li}
\end{equation}
It is noteworthy that the governing equation employed in this study, derived from the equations proposed by \cite{Deng2007Migration}, represents a pragmatic numerical approximation of wave propagation in viscoacoustic media. It is essential to acknowledge that this model assumes a frequency-independent velocity \( v \) and quality factor \( Q \) within a limited bandwidth. While this approximation neglects the subtle effects of frequency-dependent dispersion, it is a suitable tradeoff for PINNs training, as it provides a stable and differentiable residual for the loss function while maintaining sufficient accuracy for most exploration-scale applications.

In this study, the objective is to perform forward and inverse simulations for viscoacoustic wave propagation using physics-informed neural networks (PINNs). More specifically, in forward modeling, we solve the viscoacoustic wave equation by relying only on time-sparse data and physical laws, including the governing partial differential equations (PDEs) and boundary conditions. In the inversion scheme, PINNs are utilized to simultaneously reconstruct the wavefield and invert for the velocity and absorption coefficient from sparse time-domain measurements.

\section{Methodology}
In this section, we proposed a PINN architecture for both forward and inverse problems. In forward problems, a fully connected network is applied, while in inversion, we utilize a Fourier feature neural network to represent the wavefield P to mitigate the gradient vanishing problems, which lead to divergence of the network. Furthermore, considering the characteristics of seismic exploration, we apply open boundary conditions on the boundaries of the domain to eliminate the reflection.  In addition, wavefields at early timesteps by FDM are utilized as initial constraints to overcome the point-source singularity.

\subsection{PINNs}
Deep learning (DL) has garnered significant interest in seismic exploration. However, conventional DL approaches for seismic analysis typically require massive volumes of labeled training data, which are often expensive or impossible to acquire in real-world exploration. Recently, Physics-Informed Neural Networks (PINNs) have received great attention because of their ability of learning in data-scarce regimes by encoding physical laws (i.e., governing PDEs and boundary conditions) into the network constraints. 

Structurally, a physics-informed neural network (PINN) is similar to a conventional multilayer perceptron (MLP). It is composed of an input layer, several fully connected hidden layers, and an output layer. Let \( N_L \) denote the total number of layers. The mapping from the \((l-1)\)-th layer to the \(l\)-th layer can be mathematically formulated as
\begin{equation}
\mathbf{h}^{(l)} = \sigma \left( \mathbf{W}^{(l)} \mathbf{h}^{(l-1)} + \mathbf{b}^{(l)} \right), 
\quad l = 1, \ldots, N_L ,
\end{equation}
where \( \mathbf{W}^{(l)} \) and \( \mathbf{b}^{(l)} \) represent the weight matrix and bias vector of the \(l\)-th layer, respectively, and \( \sigma \) denotes a nonlinear activation function (e.g., \texttt{tanh} or \texttt{SiLU}).

The main difference between PINNs and conventional MLPs lies in the loss function. In PINNs, the loss function explicitly incorporates the governing equations as physical constraints. By minimizing this physics-informed loss function, PINNs are able to approximate solutions to partial differential equations (PDEs). For inverse problems, PINNs treat unknown physical parameters (e.g., velocity and quality factor) as trainable variables. These parameters are optimized simultaneously with the neural network weights by minimizing a composite loss function.

\subsection{Open Boundary Conditions}
In conventional grid-based methods, such as finite difference methods (FDM), Perfect Matched Layers (PML) are typically employed to eliminate boundary reflections by artificially damping waves in an extended domain. However, implementing PML requires additional auxiliary equations, which significantly complicates the loss landscape when incorporated into the optimization objective. Furthermore, PML requires the construction of extended sponge layers, inevitably leading to a larger number of collocation points and increased computational cost. Consequently, employing PML is considered computationally inefficient for the proposed PINN framework. Instead, we apply a first-order open boundary condition directly at the domain edges, which effectively allows waves to exit the computational domain without reflection.

Specifically, the formulas of first-order open boundary conditions are given by:
\begin{align}
\frac{\partial P}{\partial t} - v \frac{\partial P}{\partial x} &= 0,
\quad \text{at } x = 0 \label{eq:bc_x0} \\
\frac{\partial P}{\partial t} + v \frac{\partial P}{\partial x} &= 0,
\quad \text{at } x = L \label{eq:bc_xL} \\
\frac{\partial P}{\partial t} - v \frac{\partial P}{\partial y} &= 0,
\quad \text{at } y = 0 \label{eq:bc_y0} \\
\frac{\partial P}{\partial t} + v \frac{\partial P}{\partial y} &= 0,
\quad \text{at } y = L \label{eq:bc_yL}
\end{align}

Here, \( \{x, y, t\} \) denote the spatial and temporal coordinates, respectively. \( P \) represents the pressure wavefield, \( \text{dLen} \) is the domain length, and \( v \) is the propagation velocity. Physically, these conditions enforce one-way wave propagation behavior at the truncation edges (\( x = 0, x = L, y = 0, y = L \)), ensuring that seismic energy radiates out of the computational domain without generating reflections.

\subsection{Network architecture}
Firstly, to clarify, in order to overcome the numerical challenges posed by point-source singularities during the PINN training process, the external source term \( S \) is removed from the governing equation. Instead, the initial state of the system is constrained using early-time wavefields generated by the finite-difference method (FDM), which will be discussed later. Moreover, Eq.~\eqref{eq:viscoacoustic_wave} is reformulated by multiplying both sides by \( v^2 \) to avoid potential numerical issues associated with the \( 1 / v^2 \) term. Consequently, the governing equation employed in the proposed framework can be expressed as:
\begin{equation}
\frac{\partial^2 P}{\partial t^2} - v^2 \nabla^2 P + gv \frac{\partial P}{\partial t} = 0
\end{equation}
In the following sections, we let \( g \) denote the product of \( g v \) to simplify the notation of the loss function.

\begin{figure}
  \centering
  \includegraphics[width=0.9\linewidth]{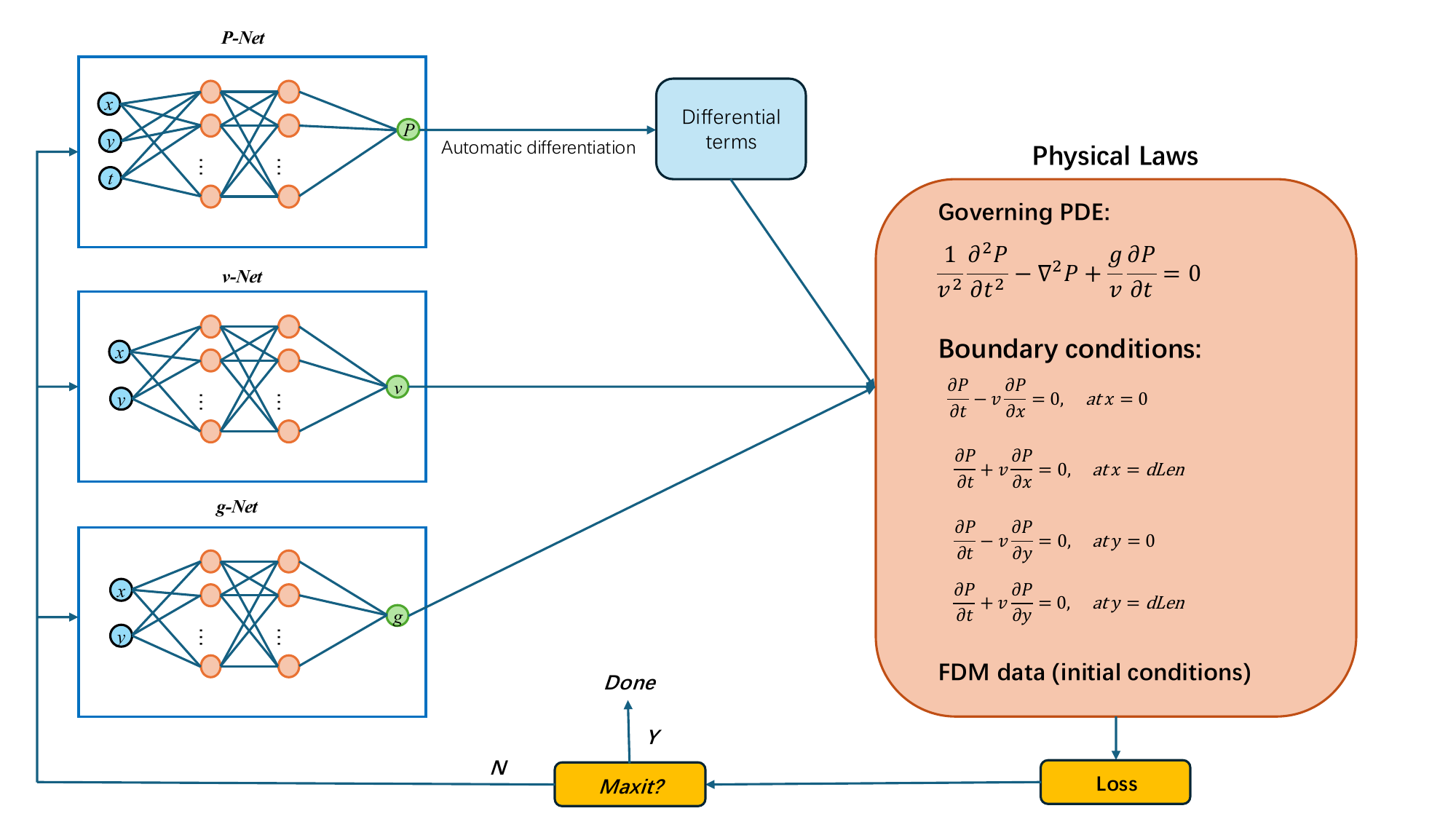}
  \caption{Overview of the proposed PINN framework for viscoacoustic wave propagation, including the network architecture and the constraint of physical laws. Three separate DNNs are used to approximate the pressure P, velocity v, and absorption coefficient g, respectively. Note that v-Net and g-Net are only activated in inversion mode. Automatic differentiation is exploited for obtaining the derivative terms and further constructing the loss function.}
  \label{fig:pinns_model}
\end{figure}

The overall architecture of the proposed PINN framework is illustrated in Fig.~\ref{fig:pinns_model}. To accommodate both forward and inverse modeling tasks, the framework adopts a modular design consisting of three distinct neural networks that approximate the pressure field \( P \) (P-Net), the velocity field \( v \) (v-Net), and the absorption coefficient \( g \) (g-Net), respectively.

For forward problems, the objective is to simulate wave propagation in a known medium. Therefore, the propagation velocity \( v \) and absorption coefficient \( g \) are fixed, implying that the v-Net and g-Net are deactivated. The network takes the mapped features \( \{x, y, t\} \) as inputs and outputs the pressure field \( P \). After obtaining \( P \), the required partial derivatives with respect to the input coordinates \( \{x, y, t\} \) are computed via automatic differentiation.

For inverse problems, the goal is to simultaneously reconstruct the wavefield and the unknown medium parameters from sparse observational data. In this case, the propagation velocity \( v \) and absorption coefficient \( g \) are treated as learnable, spatially dependent variables, such that the v-Net and g-Net are fully activated and optimized jointly with the P-Net. The P-Net takes the Fourier-mapped features \( \{x, y, t\} \) as inputs and outputs the pressure field \( P \), while the v-Net and g-Net take the spatial coordinates \( \{x, y\} \) as inputs and output \( v \) and \( g \), respectively. Notably, the predicted \( v \) and \( g \) fields are directly substituted into the governing partial differential equations (PDEs). Similar to the forward modeling case, the gradients of \( P \) with respect to the input coordinates are evaluated using automatic differentiation.

These differential terms are then substituted into the loss function constructed by concerned physical laws, including the governing equations, boundary conditions, and initial conditions. The loss function can be summarized as:
\begin{equation}
\mathcal{L} = \lambda_p \mathcal{L}_{PDE} + \lambda_b \mathcal{L}_{BC} + \lambda_d \mathcal{L}_{Data}
\end{equation}

Where \( \mathcal{L}_{\mathrm{BC}} \), \( \mathcal{L}_{\mathrm{PDE}} \), and
\( \mathcal{L}_{\mathrm{Data}} \) represent the mean squared error (MSE) of the
boundary condition mismatch, the governing equation residuals, and the data
misfit, respectively.

It is worth noting that while \( \mathcal{L}_{\mathrm{BC}} \) remains identical,
the definitions of \( \mathcal{L}_{\mathrm{PDE}} \) and
\( \mathcal{L}_{\mathrm{Data}} \) differ slightly between the forward and inverse
modes. Specifically, in the forward mode, the parameters \( v \) and \( g \) are
fixed. In this case, \( \mathcal{L}_{\mathrm{PDE}} \) minimizes the residuals of
the governing equations to solve for the wavefield \( P \) in a known medium,
while \( \mathcal{L}_{\mathrm{Data}} \) serves as an initial constraint that
enforces consistency with the initial state provided by the finite-difference
method (FDM).

In the inverse mode, the parameters \( v \) and \( g \) are treated as learnable
variables. Consequently, \( \mathcal{L}_{\mathrm{PDE}} \) jointly optimizes the
wavefield and the medium parameters, ensuring that the recovered \( v \) and
\( g \) satisfy the governing physical laws. Meanwhile,
\( \mathcal{L}_{\mathrm{Data}} \) acts as an observation constraint derived from
FDM simulations to drive the reconstruction of subsurface properties.

Here, \( \lambda_p \), \( \lambda_b \), and \( \lambda_d \) are weighting
coefficients that balance the different loss terms and are treated as
user-defined hyperparameters.

\section{Computational Experiments}
In this section, a series of computational experiments are implemented to validate the capability of the proposed PINN framework in solving forward and inverse problems for viscoacoustic wave propagation. We consider a 2D heterogeneous domain with a unified velocity or two-layered velocity model, while the viscoacoustic equation is solved within the domain. Furthermore, the inference stage is designed to evaluate the surrogate model's capability to capture the attenuation mechanism by both the variation of time-domain waveforms and frequency components. In addition, case 2 demonstrates the framework's ability to jointly solve for the forward wavefield and reconstruct the physical parameters.

\subsection{Domain Definition}
We consider the same \( 2~\mathrm{km} \times 2~\mathrm{km} \) rectangular domain for all experiments. A Ricker wavelet source located at the center of the domain is employed as the seismic source, which can be analytically expressed as:

\begin{equation}
s(t) = \left( 1 - 2(\pi f_0(t - t_0))^2 \right) \exp\left( -(\pi f_0(t - t_0))^2 \right)
\end{equation}
where the dominant frequency is set to \( f_0 = 5~\mathrm{Hz} \), and the peak time delay is set to \( t_0 = 0.2~\mathrm{s} \). In addition, the total time duration is defined as \( 950~\mathrm{ms} \).

\subsection{Case1: One-layer velocity model}
\begin{figure}
  \centering
  \includegraphics[width=\columnwidth]{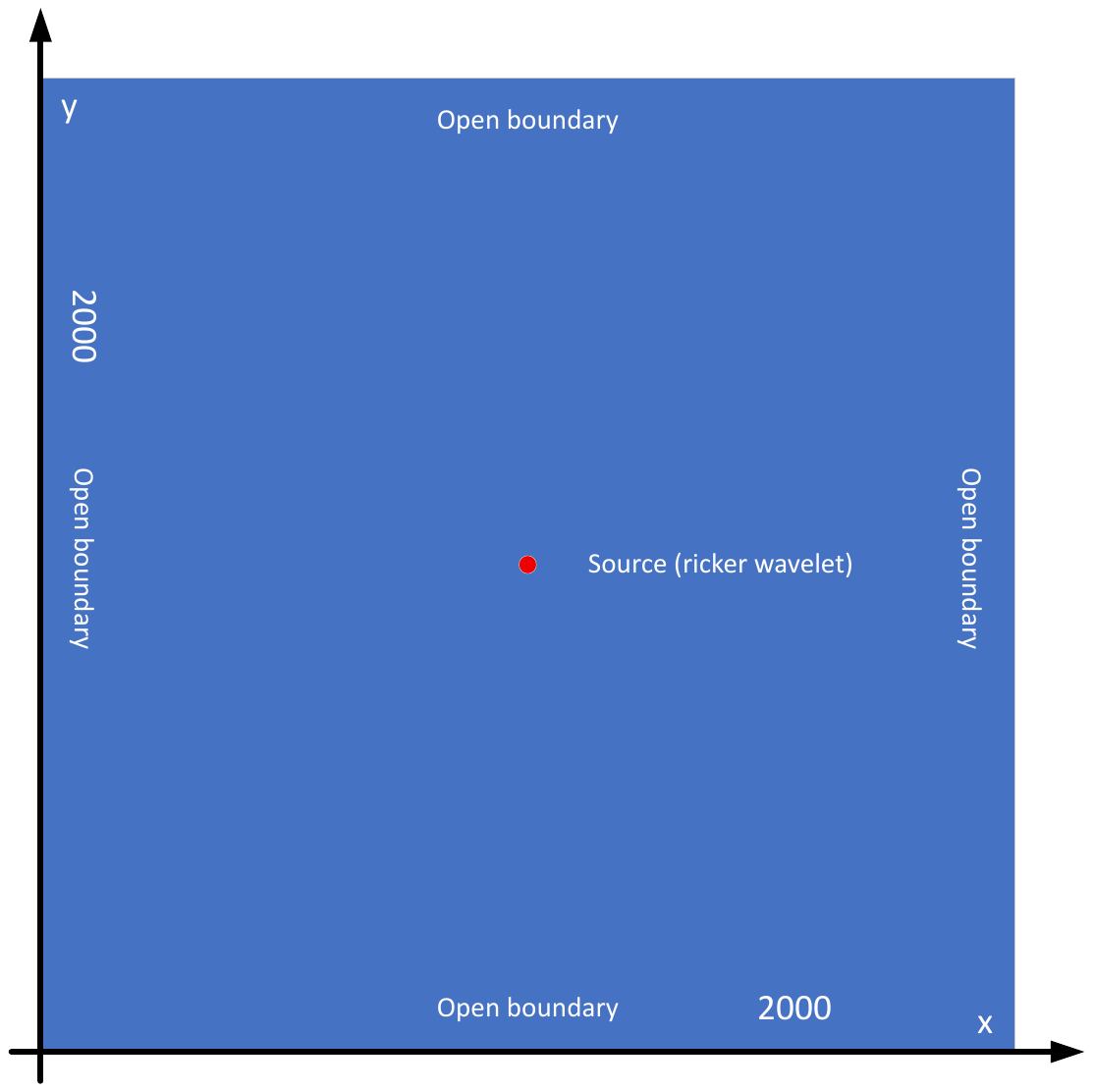}
  \caption{The spatial domain in case 1. Open boundary conditions are applied at four boundaries.}
  \label{fig:domain}
\end{figure}

In the first experiment, we consider a two-dimensional rectangular homogeneous domain with a uniform propagation velocity of \( 2~\mathrm{km/s} \). The computational domain is illustrated in Fig.~\ref{fig:domain}. While the absorption coefficient is computed using Eq.~\eqref{eq:absorption}, the quality factor \( Q \) is analytically derived from Li's empirical formula ~\eqref{eq:li}.

In this case, we sample 4,096 collocation points per batch within the interior domain to minimize the residuals of the viscoacoustic wave equation. Additionally, 1,024 points are sampled along the domain boundaries to impose the open boundary conditions, allowing seismic waves to propagate out of the truncated domain without artificial reflections. To avoid point-source singularities, wavefields at initial timesteps (150, 200, 250, and 300~ms) are used as initial constraints.

The network is trained using the Adam optimizer, and the selection of hyperparameters is detailed in the Appendix. Furthermore, the reference solutions and training datasets are generated using the finite-difference method (FDM). Specifically, we employ the open-source framework Devito \cite{Louboutin2019Devito} to numerically solve the wave equation.

\begin{figure}
  \centering
  \includegraphics[width=\columnwidth]{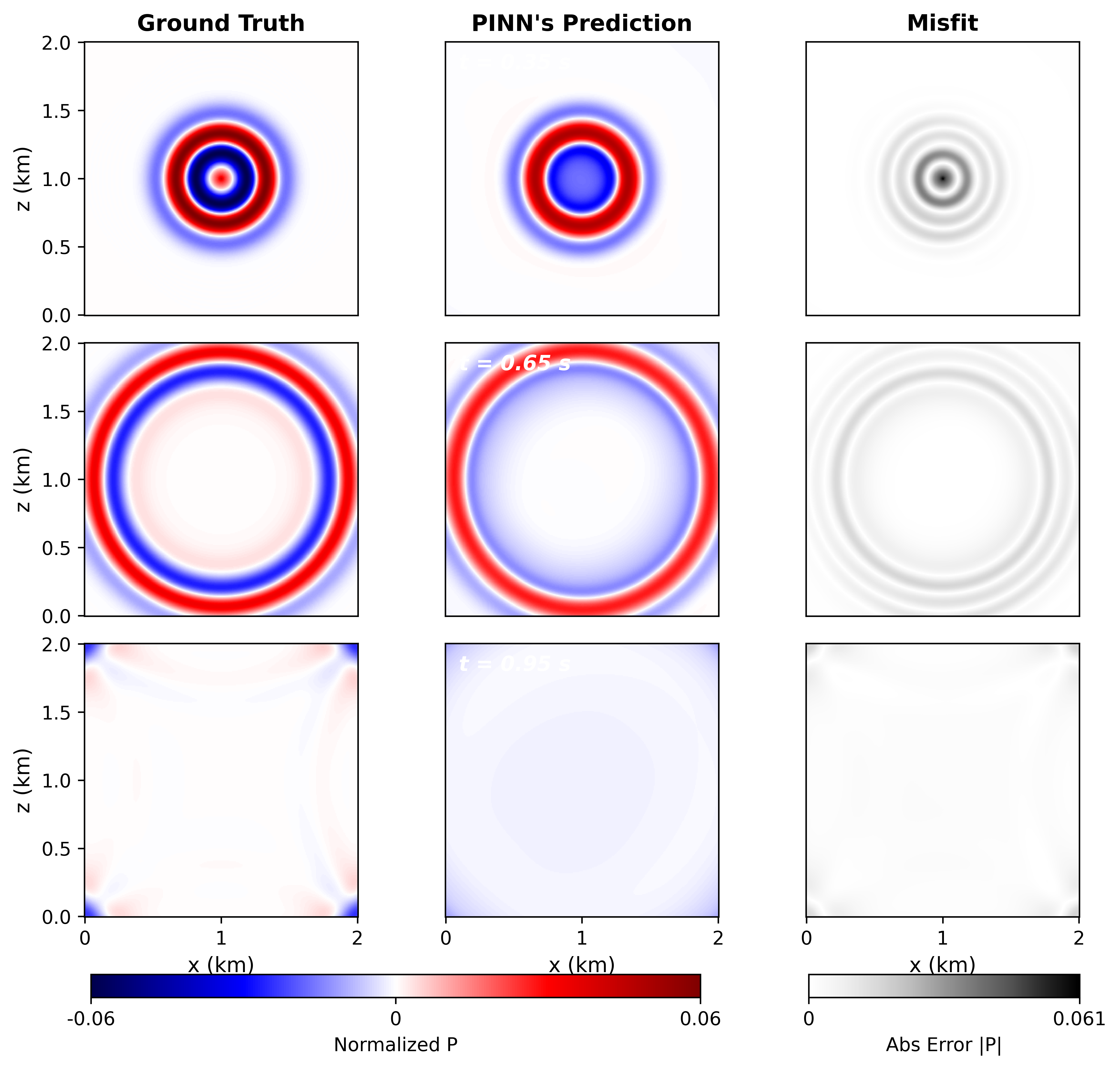}
  \caption{The results of pressure field distributions obtained from our proposed PINN and FDM in Case 1. We select three representative snapshots (i.e., t “ 0.35, 0.65, 0.95s) for comparison.}
  \label{fig:wf1}
\end{figure}

\begin{figure}
  \centering
  \includegraphics[width=\columnwidth]{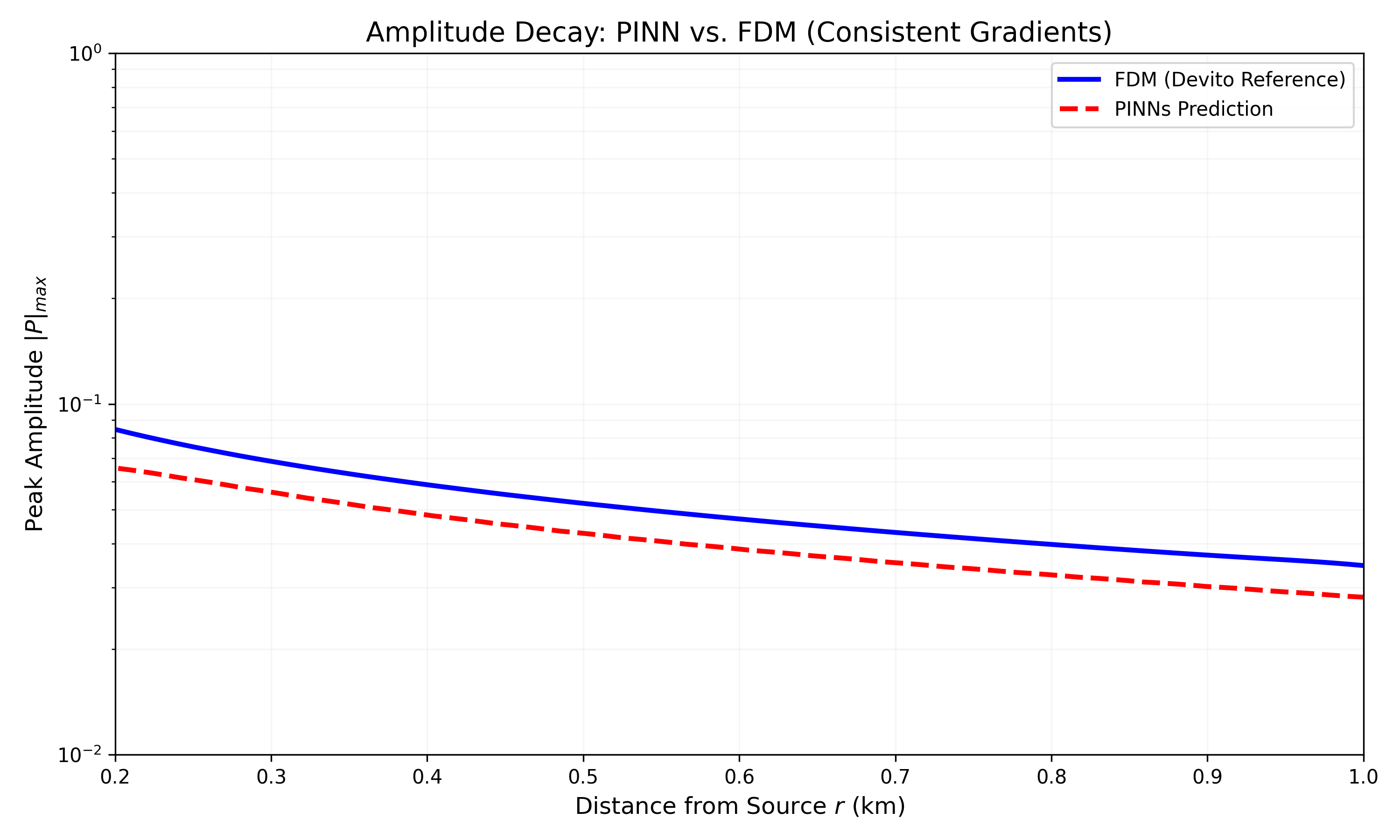}
  \caption{Comparison of amplitude decay rate for PINNs (red) and FDM(blue) in case 1}
  \label{fig:dc}
\end{figure}

\begin{figure}
  \centering
  \includegraphics[width=\columnwidth]{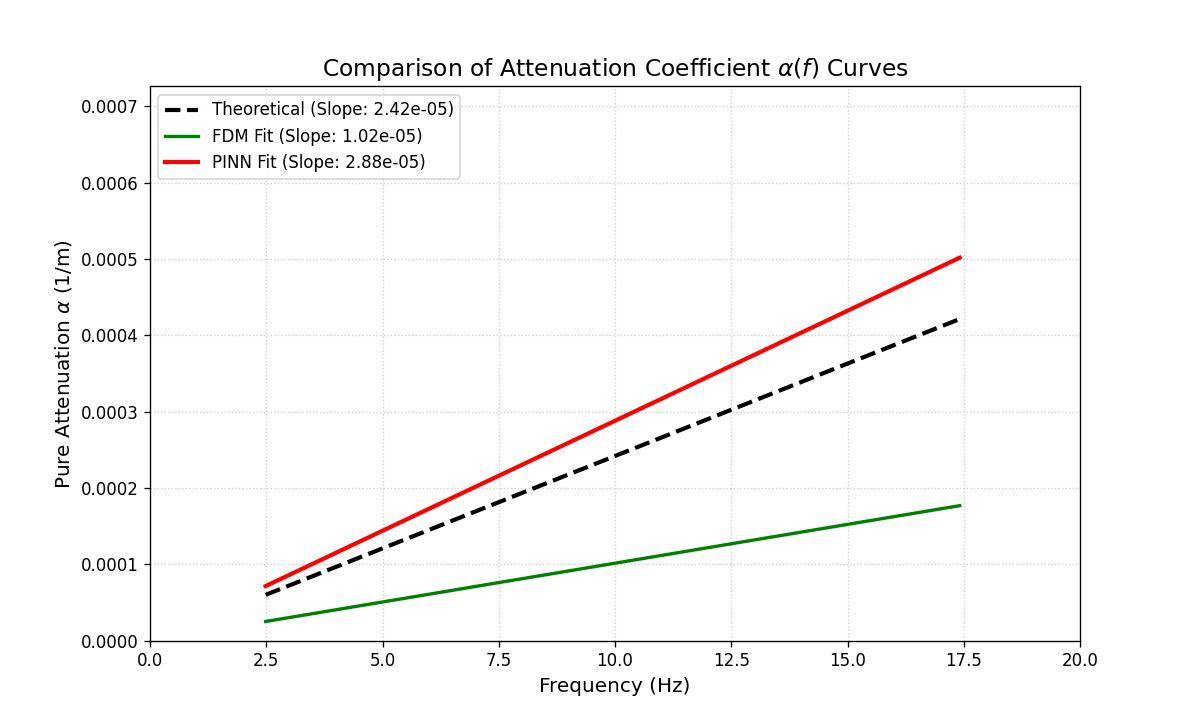}
  \caption{\(\boldsymbol{\alpha}\)--\(f\) curves by PINNs, FDM, and the analytical solution.}
  \label{fig:af}
\end{figure}

The comparative results are presented in Figs.~\ref{fig:wf1}--\ref{fig:af}. These figures demonstrate that the proposed PINN framework closely matches the ground truth. The performance of the PINN model is evaluated from three perspectives. 

Firstly, we compare the wavefield snapshots predicted by PINNs and FDM. As shown in Fig.~\ref{fig:wf1}, the snapshots produced by PINNs are in good agreement with the FDM results. Moreover, at later timesteps (e.g., 950~ms), the PINN-predicted wavefields exhibit significantly fewer boundary reflections compared to those obtained by the FDM method.

Secondly, amplitude variation is analyzed to assess whether the proposed
framework correctly reproduces the physical attenuation behavior. Specifically,
peak amplitude--distance curves extracted along the lower midline
(\( x = 1 \), \( y \in [0, 0.8] \)) are compared between the PINN and FDM solutions,as shown in Fig.~\ref{fig:dc}. 

To further quantify the amplitude behavior, in a viscoacoustic medium, the attenuation of the peak amplitude
\( A \) with respect to the propagation distance \( r \) can be expressed as
\begin{equation}
A(r, f) = A_0(f) \cdot \frac{1}{\sqrt{r}} \cdot e^{-\alpha(f)r}
\end{equation}
where \( A_0 \) denotes the initial amplitude at the source, \( r \) is the
propagation distance, and \( \alpha \) is the frequency-dependent attenuation
coefficient, defined as
\begin{equation}
\alpha(f) = \frac{\pi f}{v Q},
\label{eq:attenuation_coeff}
\end{equation}
where \( f \) is the dominant frequency of the seismic wave, \( Q \) is the
quality factor, and \( v \) is the propagation velocity. According to Li's
empirical formula \eqref{eq:li}, the quality factor \( Q \) depends only
on the velocity \( v \). Since both \( v \) and \( Q \) are assumed to be
spatially constant in this experiment, the theoretical attenuation coefficient
\( \alpha(f) \) exhibits a linear dependence on the frequency \( f \).
Consequently, the \( \alpha\)--\( f \) relationships derived from the PINN and
FDM simulations are compared against this theoretical linear trend to evaluate
their respective abilities to capture intrinsic attenuation.

Figure~\ref{fig:af} compares the attenuation coefficient
\( \alpha(f) \) estimated using different methods. The PINN-based results show
significantly better agreement with the theoretical linear attenuation trend
than those obtained using the finite-difference method (FDM). The theoretical
attenuation slope is \( 2.42 \times 10^{-5} \), while the slope derived from the
PINN results is \( 2.88 \times 10^{-5} \), corresponding to a relative error of
19.05\%. In contrast, the FDM yields a slope of \( 1.02 \times 10^{-5} \), with a
substantially larger relative error of 58.01\%. Although both methods capture
the overall frequency-dependent increase in attenuation, the PINN estimates
exhibit a more consistent linear behavior across the 2--18~Hz frequency band.
The larger deviation observed in the FDM results is attributed to numerical
dissipation and dispersion inherent to grid-based schemes, whereas the
physics-informed formulation enables PINNs to better preserve the underlying
viscoacoustic attenuation law.

Furthermore, to evaluate the robustness of the proposed PINN framework with
respect to discretization density in comparison with the traditional
finite-difference method, a comparative sensitivity analysis is conducted.
While FDM relies on discrete grid points to approximate partial derivatives,
PINNs utilize randomly sampled collocation points to evaluate the residuals of
the governing partial differential equations. To ensure a fair comparison, the
``resolution'' for both methods is defined in terms of the total number of
sampling points used per time step within the computational domain.

Specifically, a series of parallel experiments is performed by progressively
downsampling the resolution. For the FDM, the spatial grid spacing is gradually
increased (i.e., the grid is coarsened). For the PINN framework, the batch size
\( N \) of collocation points sampled within the domain is progressively
reduced. The resulting wavefields and waveforms obtained using PINNs and FDM at
different resolutions are then compared to assess the stability and robustness
of the proposed PINN approach.

\begin{figure}
  \centering
  \includegraphics[width=\columnwidth]{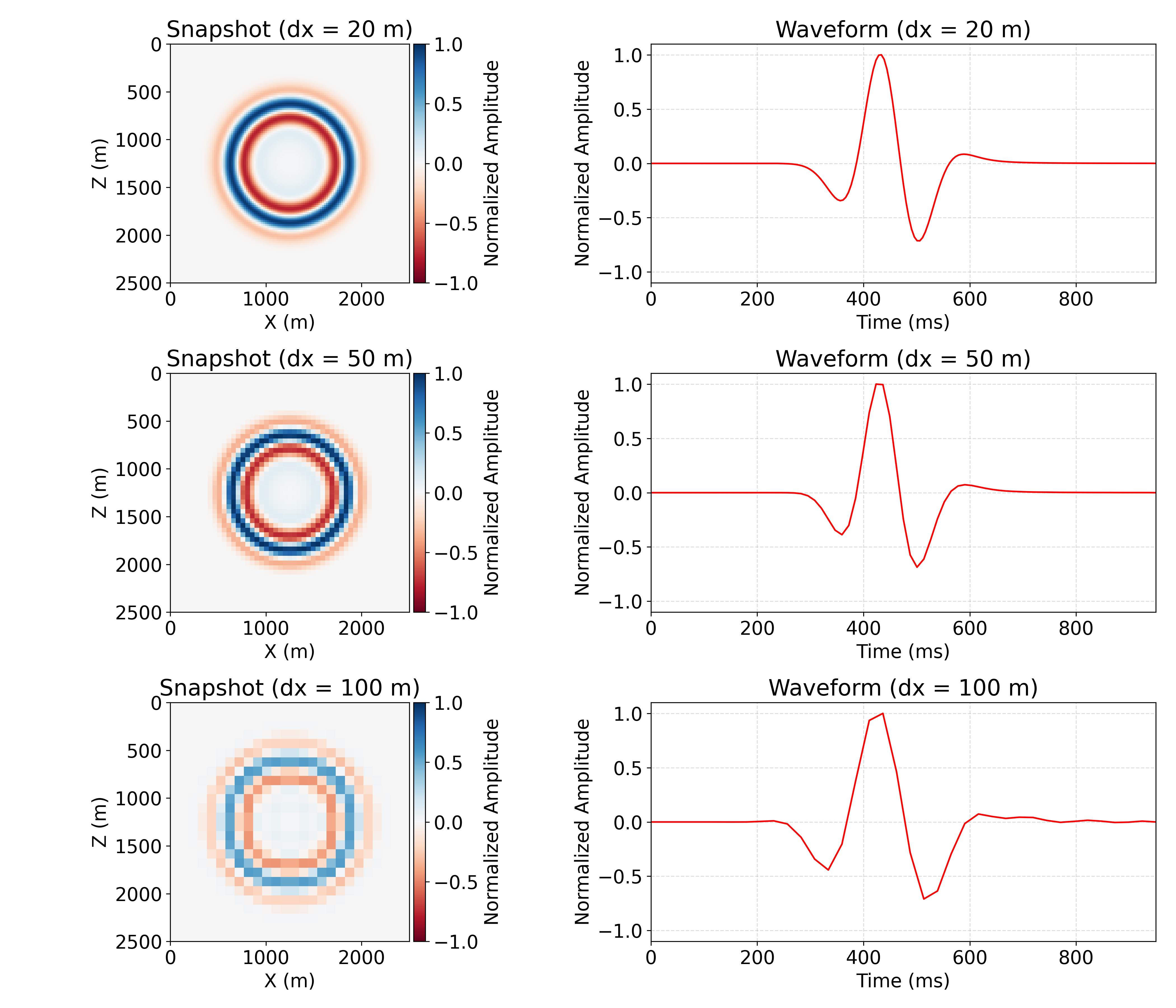}
  \caption{Wavefields and waveforms of FDM at different grid size}
  \label{fig:fs}
\end{figure}

\begin{figure}
  \centering
  \includegraphics[width=\columnwidth]{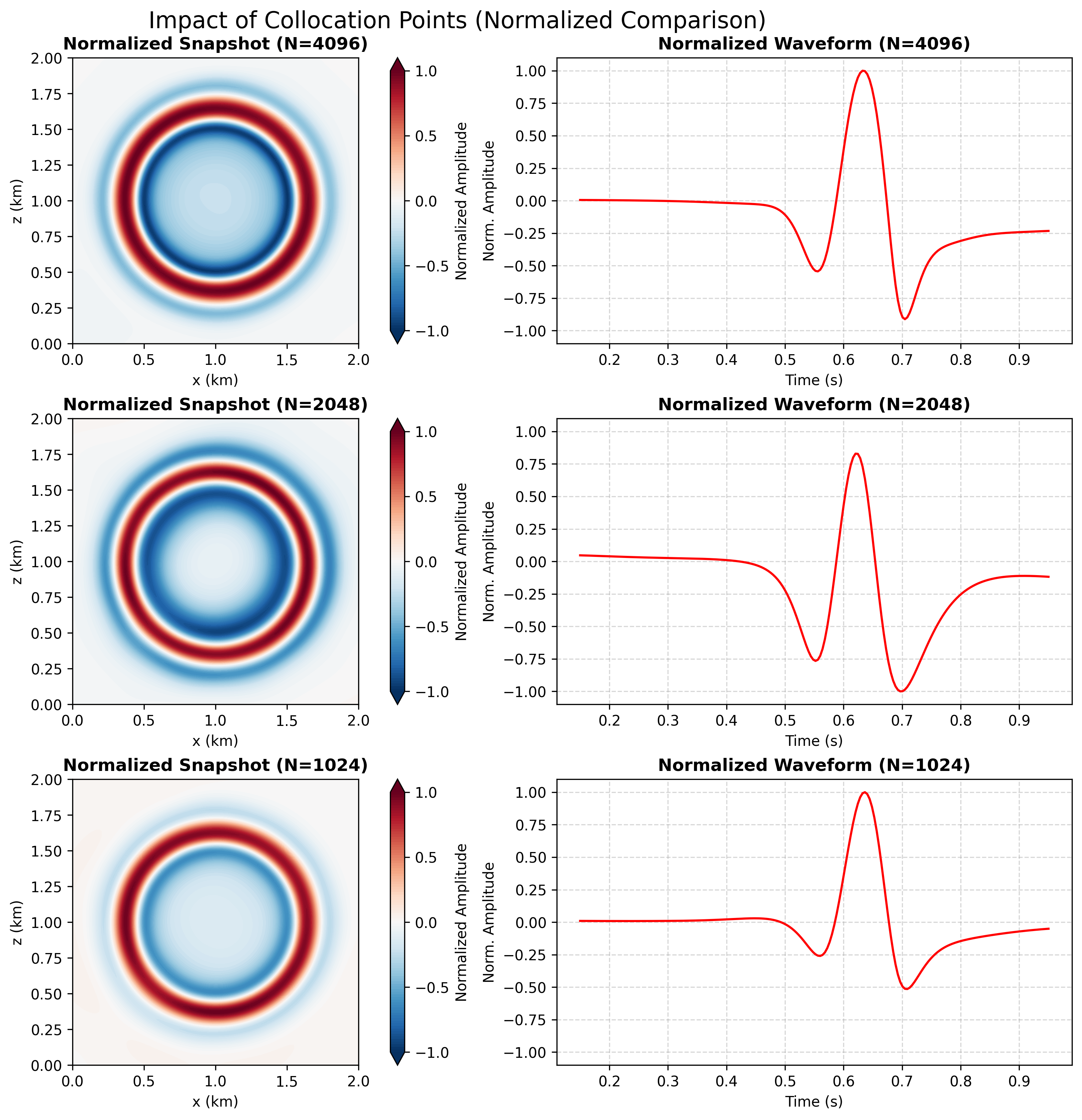}
  \caption{Wavefields and waveforms of PINNs at different collocation points}
  \label{fig:ps}
\end{figure}
Figures~\ref{fig:fs} and~\ref{fig:ps} present a comparative analysis of the PINN and FDM results under different discretization
resolutions. As shown in Fig.~\ref{fig:fs}, as the grid spacing
increases, the FDM solution exhibits pronounced numerical dispersion, manifested
by a loss of waveform fidelity and a breakdown of wavefront coherence.
Specifically, when the grid spacing is increased to
\( \Delta x = 100.0~\mathrm{m} \) (coarse-grid scenario), the FDM solution suffers
from severe structural distortion. The wavefront loses its coherent circular
shape and becomes visibly pixelated. More critically, the waveform recorded at
the monitoring point (right panel) reveals distinct spurious oscillations
trailing the main wave packet. These high-frequency tail artifacts are
non-physical and indicate that the numerical scheme fails to preserve waveform
integrity under coarse discretization.

In contrast, Fig.~\ref{fig:ps} demonstrates that the wavefields and
waveforms predicted by the PINN framework remain highly stable as the number of
collocation points decreases. Even under the extremely sparse sampling condition
of only 1,024 collocation points, the PINN model maintains a smooth and coherent
wavefield structure, effectively suppressing the spurious tail oscillations
(i.e., numerical dispersion) that typically plague FDM solutions at comparable
coarse resolutions.

\begin{figure}
  \centering
  \includegraphics[width=\columnwidth]{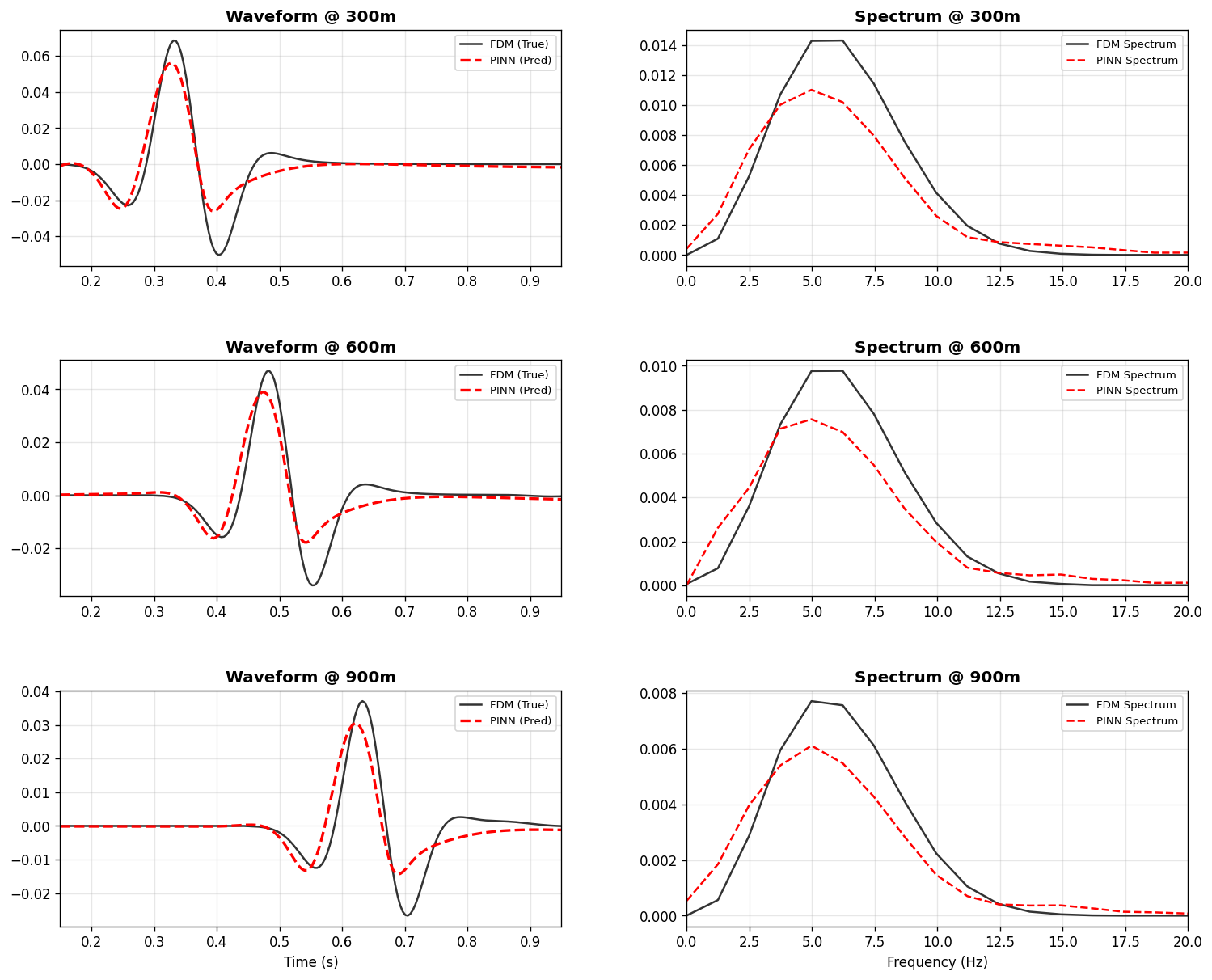}
  \caption{Waveform and frequency component comparison for PINNs (red) and FDM (black) in case 1}
  \label{fig:fr}
\end{figure}

Finally, to further validate the fidelity of the proposed PINN framework,
comprehensive comparisons of time-domain waveforms and their corresponding
frequency spectra are conducted at multiple propagation distances
(300~m, 600~m, and 900~m), as shown in Fig.~\ref{fig:fr}. While
the frequency-domain analysis confirms that the dominant frequency is accurately
preserved, the predicted waveforms (red dashed lines) exhibit a high degree of
phase synchronicity and structural similarity with the FDM benchmarks (black
solid lines). This agreement demonstrates the capability of the proposed PINN
framework to accurately capture both the temporal evolution and phase velocity
of the wavefield.

\subsection{Two-layer velocity model}
In the second experiment, we extend the application of the proposed framework
to a heterogeneous medium in order to evaluate its robustness in more complex
geological settings. In this scenario, both forward modeling and inverse
analysis are integrated to demonstrate the capability of the unified framework
to reconstruct velocity and attenuation fields from random initializations,
thereby highlighting its potential for data-efficient seismic exploration.

\begin{figure}
  \centering
  \includegraphics[width=\columnwidth]{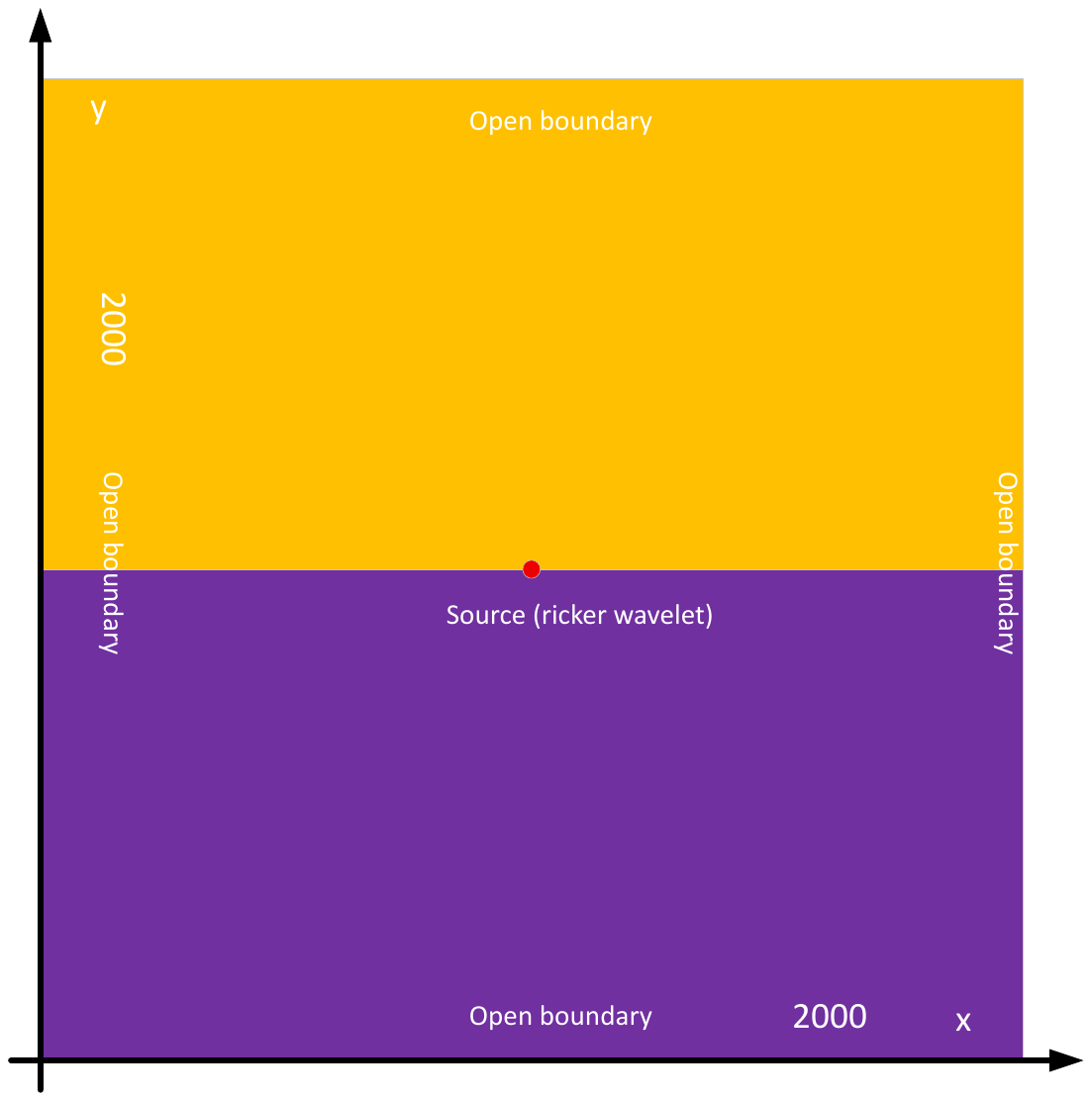}
  \caption{The spatial domain in case 2. Open boundary conditions are ap-
plied at four boundaries}
  \label{fig:d2}
\end{figure}

As illustrated in Fig.~\ref{fig:d2}, the computational domain
consists of a two-layer velocity structure. The upper layer is characterized by
a uniform propagation velocity of \( 1~\mathrm{km/s} \), while the lower layer
maintains a velocity of \( 2~\mathrm{km/s} \). This configuration introduces a
sharp impedance contrast at the horizontal interface, providing a rigorous test
of the model’s ability to enforce the viscoacoustic governing equations across
spatially varying physical properties. Similarly, the absorption coefficient
\( g \) is computed using Eq.~\eqref{eq:absorption}, and the quality factor
\( Q \) is analytically derived from Li’s empirical formula
\eqref{eq:li}.

Consistent with the first experiment, open boundary conditions are imposed on
all four boundaries to suppress artificial reflections. In addition, wavefields
at several initial timesteps are employed as constraints to circumvent the
point-source singularity.

\begin{figure}
  \centering
  \includegraphics[width=\columnwidth]{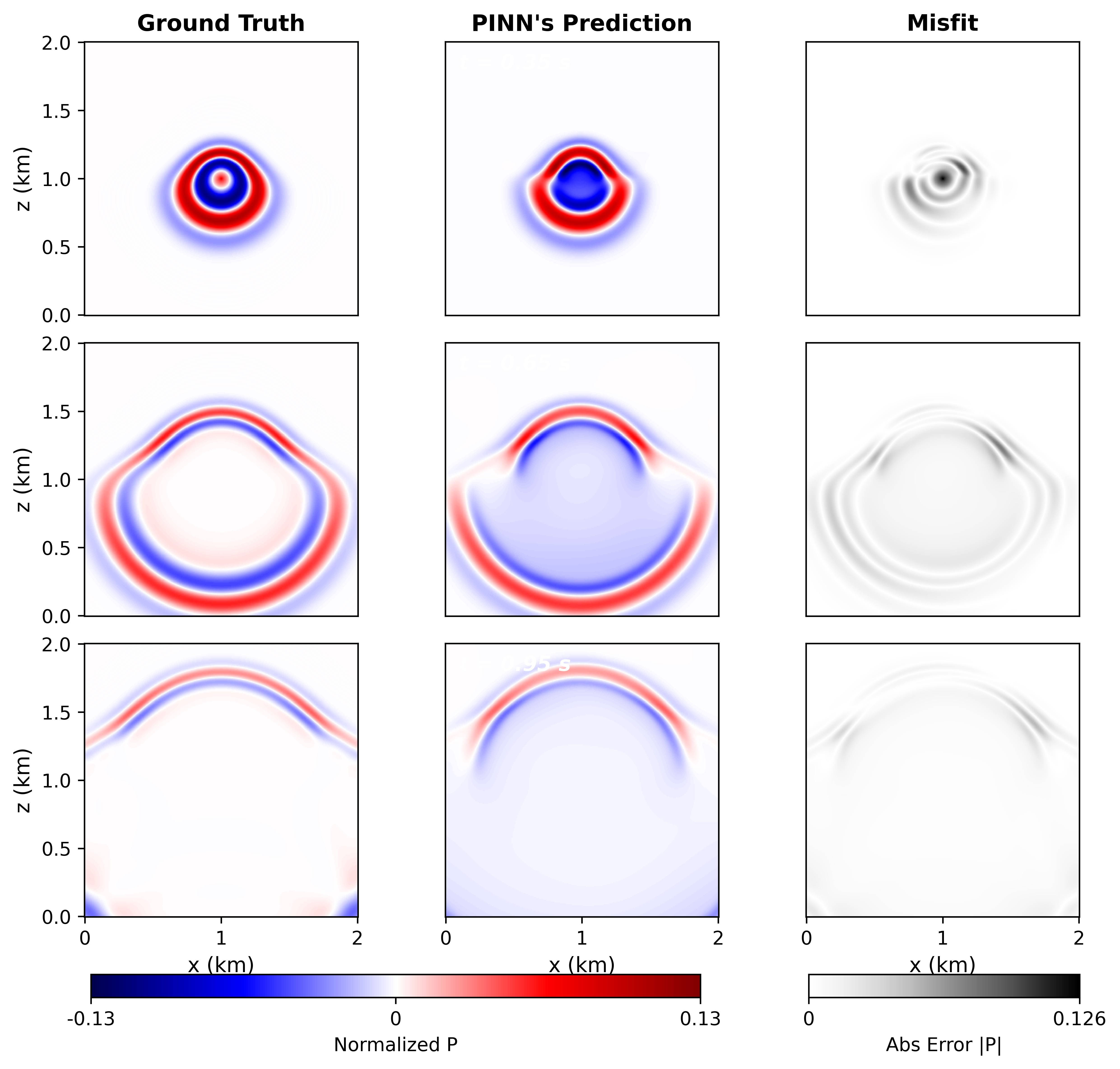}
  \caption{The results of pressure field distributions obtained from our proposed PINN and FDM in Case 2. We select three representative snapshots (i.e., t = 0.35, 0.65, 0.95s) for comparison.}
  \label{fig:wf2}
\end{figure}

\begin{figure}
  \centering
  \includegraphics[width=\columnwidth]{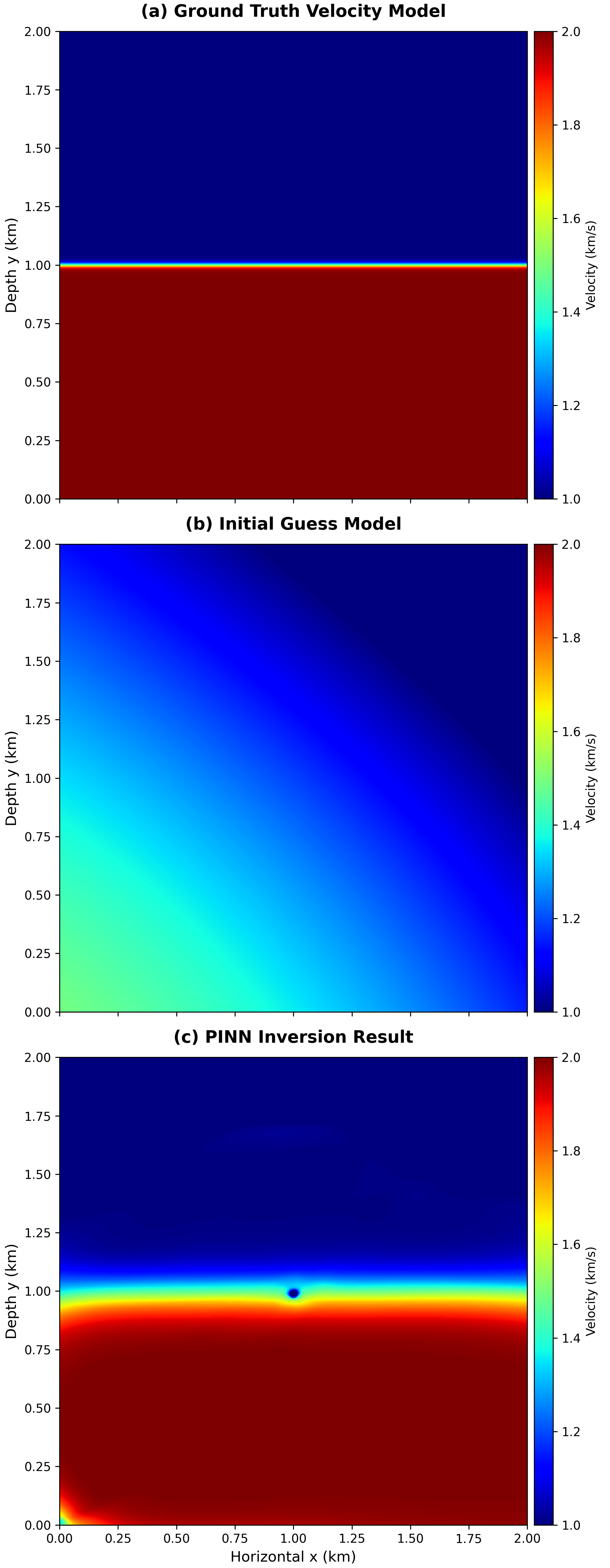}
  \caption{Ground truth, initial model, and PINN-based inversion result for velocity \(v\)}
  \label{fig:v}
\end{figure}

\begin{figure}
  \centering
  \includegraphics[width=\columnwidth]{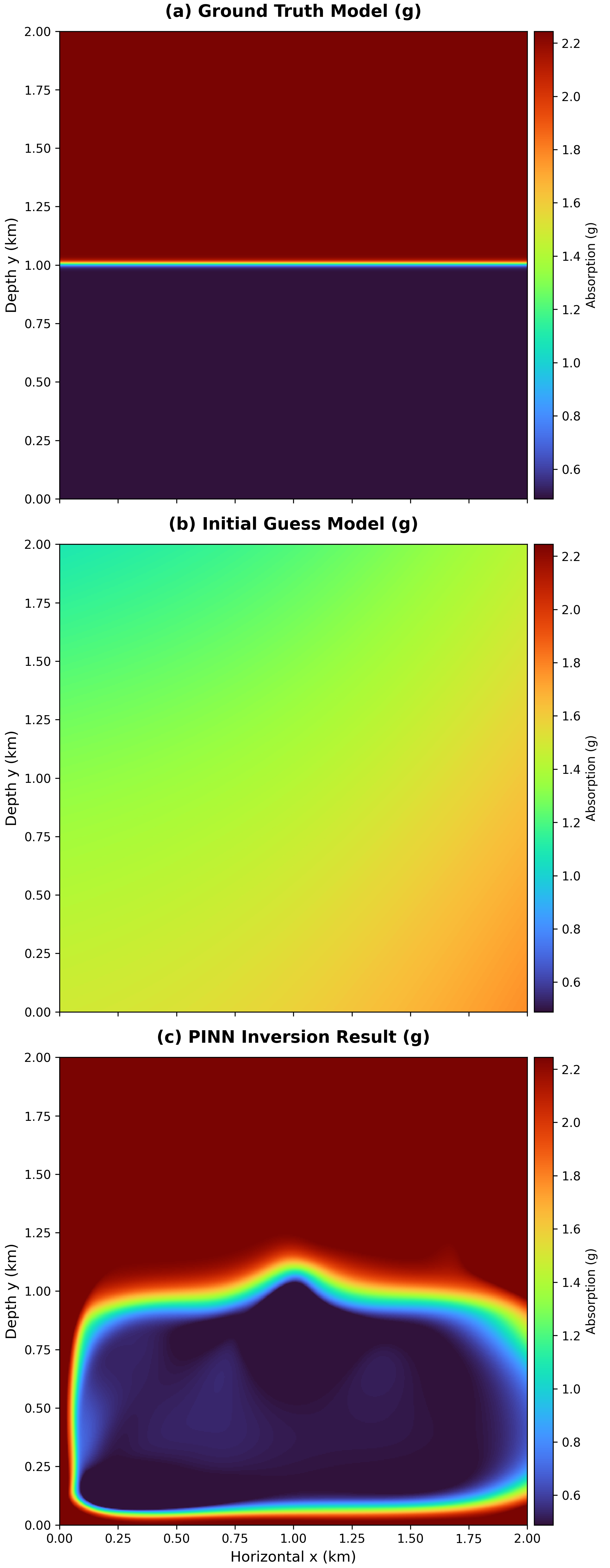}
  \caption{Ground truth, initial model, and PINN-based inversion result for absorption coefficient \(g\)}
  \label{fig:g}
\end{figure}

The comparative results between the PINN predictions and the FDM reference
solutions are presented in Figs.~\ref{fig:wf2}--\ref{fig:g}.
The performance of the proposed framework in this heterogeneous scenario is
evaluated using the following metrics.

First, similar to the first experiment, we compare the wavefield snapshots
predicted by the PINN framework with those obtained from the FDM reference.
As shown in Fig.~\ref{fig:wf2}, the PINN-predicted snapshots are in
good agreement with the FDM results (ground truth), demonstrating the
framework’s capability to resolve distinct velocity regimes in a heterogeneous
medium. Moreover, the snapshots reveal a distinctive characteristic of the
PINN-based solver. Unlike the traditional FDM solution, which exhibits a
numerically ``closed'' wavefront due to grid-based coupling across the material
interface, the PINN prediction clearly captures the physical decoupling at the
layer boundary. This behavior indicates that PINNs exhibit higher sensitivity
to sharp impedance contrasts, accurately recovering the independent propagation
characteristics of each layer without introducing artificial numerical
smoothing.

Figures~\ref{fig:v} and~\ref{fig:g} present
(a) the ground-truth models, (b) the initial guess models used by the PINN
framework, and (c) the corresponding inversion results for the velocity and
absorption coefficient fields. These results demonstrate that the proposed PINN
framework is capable of successfully recovering both the velocity and
absorption distributions from observed data, even when starting from random
initial models.

It is important to distinguish between the procedure used to generate the
ground-truth models and the inversion strategy adopted by the PINN framework.
Although the synthetic absorption coefficient \( g \) is generated from the
velocity \( v \) using Li’s empirical \( Q\)--\( v \) relationship, this
physical coupling is not imposed in either the network architecture or the loss
function. During training, the velocity and absorption fields are initialized
randomly and inverted as two independent parameters. Consequently, the accurate
reconstruction of both fields indicates that the PINN framework is capable of
disentangling kinematic effects (associated with \( v \)) and attenuation
effects (associated with \( g \)) directly from the seismic wavefield, without
relying on hard-coded rock-physics constraints.

The successful inversion demonstrates that PINNs can infer the correct physical parameters, including the correct interface and gradient changes, by minimizing the loss function under conditions of time sparsity and random initialization, suggesting that the proposed PINN approach offers a promising, data-efficient alternative to conventional inversion methods.

\section{Discussions}
This study adopts a viscoacoustic governing equation derived from
\cite{Deng2007Migration} as a pragmatic numerical approximation of viscoacoustic wave propagation. By introducing a first-order temporal derivative as an effective
damping term, the original hyperbolic wave equation is transformed into a
weakly parabolic form to emulate intrinsic attenuation effects. This
formulation assumes frequency-independent velocity and quality factor within a
limited bandwidth and therefore does not strictly satisfy causality, which in
true viscoacoustic media requires velocity dispersion. In the proposed
framework, a fixed reference frequency corresponding to the dominant source
frequency is used to compute the absorption coefficient. Although this
approximation neglects subtle frequency-dependent dispersion effects, it
successfully captures the macroscopic amplitude decay and geometric spreading
compensation required for seismic imaging and inversion at exploration scales.
More importantly, this formulation yields a numerically stable and
differentiable residual, which is particularly advantageous for training
physics-informed neural networks.

A critical finding of this study is the fundamentally different failure
behaviors exhibited by the finite-difference method (FDM) and PINNs under
low-resolution conditions. Sensitivity analyses indicate that FDM suffers from
a threshold-type failure: once the grid spacing exceeds the numerical stability
limit (e.g., \( \Delta x = 100~\mathrm{m} \)), severe numerical dispersion
emerges, characterized by spurious oscillatory tails and pronounced phase
distortion. In contrast, the proposed PINN framework demonstrates a graceful
degradation behavior. Even under extremely sparse sampling conditions, such as
using only 1,024 collocation points, the solution remains stable, albeit with
reduced accuracy. This robustness primarily arises from the fundamentally
different solution paradigms of PINNs and FDM. While FDM relies on local,
grid-based time marching and is therefore highly sensitive to discretization
thresholds, PINNs formulate wave propagation as a global optimization problem
constrained by the governing equations. Stochastic resampling of collocation
points further enhances this property by improving coverage of the
spatiotemporal domain during training.

The successful recovery of both velocity and attenuation fields from stochastic
initial models further demonstrates that PINNs can effectively solve inverse
problems by treating physical parameters as learnable variables. Unlike
traditional forward modeling approaches with fixed model parameters, the PINN
framework simultaneously optimizes the neural network weights and physical
parameters (velocity and attenuation) through gradient-based optimization,
thereby enabling unified forward and inverse modeling within a single
framework.

Nevertheless, several limitations of the PINN framework should be
acknowledged. First, the spectral bias inherent to deep neural networks leads to
preferential learning of low-frequency components, resulting in diminished
high-frequency content compared to FDM solutions. Second, for simple forward
modeling tasks, PINNs are computationally less efficient than classical FDM
solvers, as iterative optimization is required even for homogeneous models.
Finally, the performance of PINNs remains sensitive to hyperparameter selection,
particularly the weighting of different loss components.

Future work will focus on adaptive loss-weighting strategies and automated
hyperparameter optimization to improve robustness across different physical
regimes. In addition, advanced network architectures incorporating attention
mechanisms, such as Transformer-based models, may help mitigate spectral bias
and better capture long-range dependencies in seismic wavefields.

\section{Conclusion}
This study presents a physics-informed neural network (PINN) framework for viscoacoustic wave modeling and inversion. By incorporating an effective damping formulation to approximate intrinsic attenuation, the proposed approach enables a stable representation of viscoacoustic wave propagation. Numerical experiments demonstrate that both velocity and absorption fields can be successfully recovered from random initial models, highlighting the inversion capability of PINNs by treating physical parameters as learnable variables. Another key observation is the fundamentally different behavior exhibited by PINNs and finite-difference methods under coarse discretization or data-scarce conditions. While finite-difference solvers suffer from abrupt failure due to numerical dispersion once stability limits are exceeded, the PINN framework exhibits a gradual degradation in accuracy while maintaining solution stability. These results suggest that PINNs provide a flexible and robust alternative for viscoacoustic seismic modeling and inversion, particularly in scenarios where dense spatial or temporal sampling is impractical. 

\section*{Acknowledgements} Liang~CH gratefully acknowledges the continuous guidance and support of his supervisor, Professor Jun~Matsushima, as well as the revisions and geophysical insights provided by Dr.~Peng~XL. He also sincerely thanks the co-supervisor, Professor Katsunori~Mizuno, for his specialized and insightful suggestions. In addition, special thanks are extended to former visiting PhD student Lou~JS and all colleagues in the laboratory for their constructive discussions related to this thesis.

\bibliographystyle{gji}
\bibliography{refs}

\end{document}